\documentstyle[sprocl]{article}

\input epsf.sty

\newcommand{\insertfig}[2]{\mbox{\epsfxsize=#1cm \epsfbox{#2.eps}}}
\newcommand{\g}{{\sl g}}
\newcommand{\ft}[2]{{\textstyle\frac{#1}{#2}}}

\def\be{\begin{equation}}
\def\ee{\end{equation}}
\def\bea{\begin{eqnarray}}
\def\eea{\end{eqnarray}}
\begin{document}

%%%%%%%%%%%%%%%%%%%%%%%%%%%%%%%%%%%%%%%%%%%%%%%%%%%%%%%%%%%%%%%%%%%%%%%%
\begin{titlepage}

\centerline{\large \bf Integrability of twist-three evolution equations
                       in QCD.}

\vspace{10mm}

\centerline{\bf A.V. Belitsky}

\vspace{10mm}

\centerline{\it C.N. Yang Institute for Theoretical Physics}
\centerline{\it State University of New York at Stony Brook}
\centerline{\it NY 11794-3800, Stony Brook, USA}

\vspace{3mm}

\centerline{\it Institut f\"ur Theoretische Physik,
                Universit\"at Regensburg}
\centerline{\it D-93040 Regensburg, Germany}

\vspace{30mm}

\centerline{\bf Abstract}

\vspace{0.8cm}

We describe a recent progress in finding solutions to three-particle
evolution equations at leading order in the QCD coupling constant for
multiparton correlation functions based on the integrability of
corresponding interaction Hamiltonians.

\vspace{30mm}

\centerline{\it Talk given at the}
\centerline{\it 4th Workshop on Continuous Advances in QCD}
\centerline{\it Minneapolis, May 12-14, 2000}

\end{titlepage}
%%%%%%%%%%%%%%%%%%%%%%%%%%%%%%%%%%%%%%%%%%%%%%%%%%%%%%%%%%%%%%%%%%%%%%%%

\title{INTEGRABILITY OF TWIST-THREE EVOLUTION EQUATIONS IN QCD}

\author{A. V. BELITSKY}

\address{C.N.\ Yang Institute for Theoretical Physics, \\
         State University of New York at Stony Brook, \\
         NY 11794-3840, Stony Brook, USA}

\address{Institut f\"ur Theoretische Physik, Universit\"at Regensburg \\
         D-93040 Regensburg, Germany}


\maketitle\abstracts{
We describe a recent progress in finding solutions to three-particle
evolution equations at leading order in the QCD coupling constant for
multiparton correlation functions based on the integrability of
corresponding interaction Hamiltonians. }

%%%%%%%%%%%%%%%%%%%%%%%%%%%%%%%%%%%%%%%%%%%%%%%%%%%%%%%%%%%%%%%%%%%%%
\section{Multiparton correlations in hard scattering.}
\label{multi}
%%%%%%%%%%%%%%%%%%%%%%%%%%%%%%%%%%%%%%%%%%%%%%%%%%%%%%%%%%%%%%%%%%%%%

\begin{figure}[t]
\begin{center}
\hspace{0cm}
\mbox{
\begin{picture}(0,25)(100,0)
\put(-40,0){\insertfig{10}{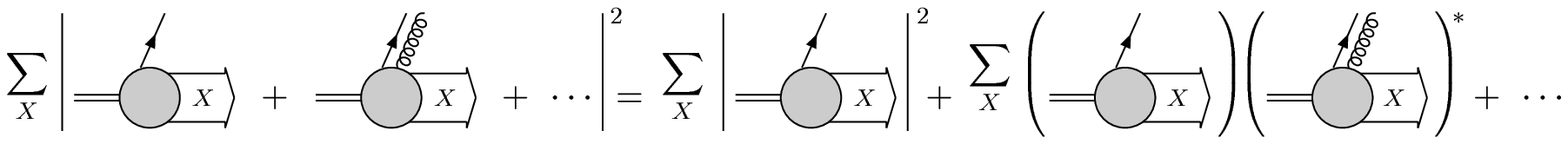}}
\end{picture}
}
\end{center}
\caption{\label{hard-interf} Hadronic structure as a series in a number of
Fock components participating in a hard rescattering. The first term on
the r.h.s.\ of the equality stands for leading twist-two contributions.
The second one (plus its complex conjugate), to the twist-three effects,
etc.}
\end{figure}

A typical lepton-hadron cross section at high momentum transfer $Q$
is given by a series in the latter
\begin{equation}
\label{twist-exp}
\sigma = \sum_{\tau = 2}^{\infty}
\left( \frac{\Lambda}{Q} \right)^{\tau - 2}
\int \{ dx_\tau \} C \left.\left( x, \{ x_\tau \} \right| \alpha_s \right)
f \left( \{ x_\tau \}, Q^2 \right) ,
\end{equation}
where $\tau$ stands for the twist of contributing operators which
parametrize the physics at soft scales. The first term in the expansion
(\ref{twist-exp}) corresponds to the Feynman model picture. Consequent
terms, standing for multi-parton correlations in hadron, reveal QCD dynamics
not encoded into the conventional parton densities. They manifest genuine
quantum mechanical effects being an interference of hadron wave functions
with different number of particles, see Fig.\ \ref{hard-interf}. Twist-three
correlations are unique among other higher-twist effects by their feature
to appear as a leading effect in certain spin asymmetries. The most
familiar examples are the transverse spin structure function $g_2$
measured in deep inelastic scattering. If a cross section is measured
in a rather wide range of hard momentum $Q$, the logarithmic scaling
violation in the functions $f$ in Eq.\ (\ref{twist-exp}) becomes an issue
since hadron substructure is seen by a probe with different resolutions.
In QCD it is formalized by means of renormalization group
equations with kernels which are derived by standard methods of QCD
perturbation theory (see e.g.\ \cite{BukKurLip83,Bel97,KodTan98}). The
specifics of twist-three sector as compared to the twist-two case is that
the afore mentioned structure functions being defined as Fourier transform
($FT$) of two-particle hadron-hadron matrix elements $f \sim FT \langle h
| \phi_1 \phi_2 | h \rangle$ receive contributions from three-parton
quark-gluon correlators \cite{BukKurLip83,Bel97,KodTan98} plus a
kinematical piece from twist-two operators going under the name of
Wandzura-Wilczek part. The original operators, not possessing a definite
twist, mix with two- as well as three-particle operators and thus evolution
equation will have an extremely complicated form. Resolving this
complication in favour of an independent study of the renormalization
group evolutions of separate twist components one finds an autonomous
equations for two- and three-particle sector, respectively. The twist-two
sector is simple and is well-studied in the literature. It is the latter
which is the subject of this presentation. In spite of simplifications
due to reduction alluded to before, anyway, one ends up with a very
complicated problem of working out the mixing of three-particle local
operators.

%%%%%%%%%%%%%%%%%%%%%%%%%%%%%%%%%%%%%%%%%%%%%%%%%%%%%%%%%%%%%%%%%%%%%
%                           Figure 1                                %
%%%%%%%%%%%%%%%%%%%%%%%%%%%%%%%%%%%%%%%%%%%%%%%%%%%%%%%%%%%%%%%%%%%%%
%%%%%%%%%%%%%%%%%%%%%%%%%%%%%%%%%%%%%%%%%%%%%%%%%%%%%%%%%%%%%%%%%%%%%

%%%%%%%%%%%%%%%%%%%%%%%%%%%%%%%%%%%%%%%%%%%%%%%%%%%%%%%%%%%%%%%%%%%%%
\section{Leading order evolution.}
\label{evolution}
%%%%%%%%%%%%%%%%%%%%%%%%%%%%%%%%%%%%%%%%%%%%%%%%%%%%%%%%%%%%%%%%%%%%%

As we have stated above reducing the twist-three two-particle operators
to the three-particle ones $\phi_1 \phi_2 \phi_3$ allows to easily find
corresponding evolution equation since the latter operators fall into a
class of the so-called quasi-partonic \cite{BukFroLipKur85} ones and thus
have a bunch of remarkable properties which simplifies their renormalization
properties. Namely, at leading order in the QCD coupling constant the total
kernel $\mbox{\boldmath$K$}_{123}$ reduces to the sum of conventional
pair-wise twist-two non-forward kernels $K_{ab}$ which includes the
momentum conservation delta function $\delta (k_a + k_b - k'_a - k'_b)$.
Thus, the leading order evolution equation is of the form
\begin{equation}
\label{EvolEq}
\frac{d}{d\ln Q^2} F_{123}
= - \frac{\alpha_s}{2 \pi} \mbox{\boldmath$K$}_{123} \ast F_{123} ,
\qquad\mbox{with}\qquad
\mbox{\boldmath$K$}_{123} = K_{12} + K_{23} + K_{13} .
\end{equation}
Here $F_{123} \equiv F (x_1, x_2, x_3) $ and $\ast \equiv \int dx_1 dx_2
dx_3 \delta \left( x_1 + x_2 + x_3 - \eta \right)$ and variable $\eta$
which encodes the skewedness of the matrix element: $\eta = 0$ for the
forward scattering and $\eta = 1$ for totally exclusive kinematics. The
solution to Eq.\ (\ref{EvolEq}) is expected to be of the form
\begin{equation}
F_{123} ( Q^2 )
= \sum_{\{ \alpha \}}
{\mit\Psi}_{\{ \alpha \} \, 123}
\left(
\frac{\alpha_s ( Q_0^2 )}{\alpha_s ( Q^2 )}
\right)^{f_{(c)} {\cal E}_{\{ \alpha \}} / \beta_0}
\langle\langle {\cal F}_{\{ \alpha \}} ( Q_0^2 ) \rangle\rangle ,
\end{equation}
where ${\cal E}_{\{\alpha\}}$ are the eigenvalues of evolution kernels
(and $f_{(c)}$ is an extracted colour factor) and ${\mit\Psi}_{\{ \alpha
\}} (x, x')$ are the corresponding eigenfunctions pa\-ra\-met\-ri\-zed
by a set quantum numbers $\{ \alpha \}$. As usual $\beta_0 = \frac{4}{3}
T_F N_f - \frac{11}{3} C_A$ is the leading term of the QCD
$\beta$-function and $\langle\langle {\cal F}_{\{ \alpha \}} ( Q_0^2 )
\rangle\rangle$ stand for reduced matrix elements of local operators at
a low normalization point $Q_0$. Therefore the evolution equation
(\ref{EvolEq}) can be reduced to a stationary Schr\"odinger equation
\begin{equation}
\label{ThreePartProblem}
\mbox{\boldmath$K$}_{123} \ast {\mit\Psi}_{\{ \alpha \} \, 123}
= f_{(c)} {\cal E}_{\{ \alpha \}} {\mit\Psi}_{\{ \alpha \} \, 123} .
\end{equation}
Simplifications in diagonalization of this equation occur due to use of
QCD symmetries which we now use in turn.

%%%%%%%%%%%%%%%%%%%%%%%%%%%%%%%%%%%%%%%%%%%%%%%%%%%%%%%%%%%%%%%%%%%%%
\section{Conformal symmetry.}
\label{conformal}
%%%%%%%%%%%%%%%%%%%%%%%%%%%%%%%%%%%%%%%%%%%%%%%%%%%%%%%%%%%%%%%%%%%%%

The classical QCD Lagrangian enjoys the property of invariance under the
15-parameter group of conformal transformations which consists of the
Lorentz group, ${\cal P}_\mu$ and ${\cal M}_{\mu\nu}$, and generators
of dilatation ${\cal D}$ and special conformal ${\cal K_\mu}$
transformations \cite{MacSal69}. At quantum level the trace of the energy
momentum tensor develops an anomaly ${\mit\Theta}_{\mu\mu} \neq 0$ and
the generators ${\cal D}$ and ${\cal K}_\mu$ cease to be symmetries of
the theory. The invariance of Lagrangian under a given transformation
imposes severe restrictions on the form of eigenfunctions of composite
operators entering the evolution. For illustration purposes let us consider
the situation with two-particle operators since they enter as a building
block in $(ab)$-subchannel for our general consideration of three-particle
problem. It was known for a long time that these are conformal operators
\cite{EfrRad79} which diagonalize the leading order renormalization group
equation for local operators with total derivatives, e.g.\ $\phi_b
(\partial_a + \partial_b)_+^k (\partial_a - \partial_b)^j_+ \phi_a$, where
the $+$-subscript stands for projection with a light-like vector $n_\mu$,
i.e.\ $v_+ = v_\mu n_\mu$. Let us demonstrate it explicitly. The conformal
operators have the form
\begin{equation}
\label{ConfOper}
{\cal O}_{jk}
= \phi_b
(i \partial_a + i \partial_b)^k_+ P^{(\nu_b, \nu_a)}_j
\left(
(\partial_a - \partial_b)_+ / (\partial_a + \partial_b)_+
\right) \phi_a ,
\end{equation}
where $P^{(\nu_b, \nu_a)}_j$ are Jacobi polynomials with $\nu_a = d_a + s_a
- 1$, and $d_a$ and $s_a$ being dimension and spin of a field $\phi_a$.
Since we deal with operators on the light-cone it is enough to consider
a collinear sub-group of the conformal group, which consists of
projections ${\cal P}_+ \equiv {\cal P}_\mu n_\mu$, ${\cal M}_{-+} =
{\cal M}_{\mu\nu} n^\star_\mu n_\nu$, ${\cal D}$ and ${\cal K}_- =
{\cal K}_\mu n^\star_\mu$, forming an $so(2, 1)$ algebra, $[{\cal J}^3 ,
{\cal J}^\pm]_- = \pm {\cal J}^\pm$, $[{\cal J}^+ , {\cal J}^-]_- = - 2
{\cal J}^3$, defined by generators of the momentum ${\cal J}^+ = i
{\cal P}_+$, special conformal transformation ${\cal J}^- = \frac{i}{2}
{\cal K}_-$ and a particular combination of dilatation and angular
momentum ${\cal J}^3 = \frac{i}{2} ({\cal D} + {\cal M}_{-+})$. The
Casimir operator is $\mbox{\boldmath${\cal J}$}^2 = {\cal J}^3 ( {\cal J}^3
- 1 ) - {\cal J}^+ {\cal J}^-$. Simple calculation shows that the operators
(\ref{ConfOper}) transform covariantly under dilatation and special
conformal transformation, namely,
\begin{eqnarray}
\label{confVaria}
&&i [ {\cal O}_{jl}, {\cal P}_+ ]_-
= i\, {\cal O}_{jl + 1} ,
\qquad\quad
i [ {\cal O}_{jl}, {\cal M}_{-+} ]_-
= -( l + s_a + s_b ) {\cal O}_{jl} , \nonumber\\
\label{confVaria-1}
&&i [{\cal O}_{jl}, {\cal D} ]_-
= - ( l + d_a + d_b ) {\cal O}_{jl},
\qquad
i [ {\cal O}_{jl}, {\cal K}_- ]_-
= i\, a (j,l) {\cal O}_{jl-1} ,
\label{MatrixA}
\end{eqnarray}
with $a_{jk}(l) = \delta_{jk} \cdot a(j,l)$ and $a (j, l) = 2
(j - l)( j + l + \nu_a  + \nu_b )$. From these equations one easily
sees the meaning for ${\cal J}^\pm$ to be step-up and -down operators
as ${\cal P}_+$ adds one unit of spin to the operator in a conformal
tower while ${\cal K}_-$ reduces it by one. These operators possess
conformal spin $J_{ab} = j + \frac{1}{2}(\nu_a + \nu_b + 2 )$,
$[\mbox{\boldmath${\cal J}$}^2_{ab}, {\cal O}_{jk}]_- = J_{ab}(J_{ab} - 1)
{\cal O}_{jk}$, spin $l + s_a + s_b$ and scale dimension $l + d_a + d_b$.
Thus conformal operators form an irreducible infinite dimensional
representation of the collinear conformal group spanned by bilocal
operators.

The renormalization group equation for conformal operators have the
form
\begin{equation}
\label{RGequation}
\frac{d}{d \ln Q} {\cal O}_{jl}
= - \sum_{k = 0}^{j} \gamma_{jk} {\cal O}_{kl},
\end{equation}
with triangular matrix of anomalous dimensions $\gamma_{jk}$, $j \geq k$.
To find a constraint on the form of the anomalous dimension matrix of
${\cal O}_{jk}$ stemmed from the special conformal symmetry, we use Ward
identities for ${\cal D}$ and ${\cal K}_-$ transformations. Since the
dilatation WI is equivalent to the renormalization group equation
(\ref{RGequation}) we get an equation on $\gamma_{jk}$. Explicit analysis
gives \cite{BelMul98}
\begin{equation}
\gamma_{jj'} \, a_{j'k} (l) - a_{jj'} (l) \, \gamma_{j'k} = 0 ,
\end{equation}
at leading order in coupling constant, $\gamma \sim {\cal O} (\alpha_s)$.
Since the matrix $a$ in (\ref{MatrixA}) is diagonal it forces $\gamma$
to be diagonal as well at lowest order in strong constant, $\gamma_{jk}
= \gamma_j \delta_{jk}$. Beyond this order conformal symmetry is
violated and the diagonal matrix $a$ is promoted to a non-diagonal one
\cite{BelMul98} $a_{jk} \to a_{jk} + \gamma^c + 2 \ft{\beta}{\g} \, b$,
with new objects which are special conformal anomaly $\gamma^c$ appearing
from the renormalization of the trace anomaly and a conformal operator
${\cal O}_{jl} \int\! dx\, x_- {\mit\Theta}_{\mu\mu} \propto \sum_{k =
0}^{j} \gamma^c_{jk} {\cal O}_{kl}$, and an $\alpha_s$-independent matrix
$b$. These extra terms induce the non-diagonal part of the anomalous
dimensions beyond leading order.

Therefore, the net product of analysis is that due to preservation of
conformal invariance for the leading order anomalous dimensions (pair-wise
kernels of the previous section \ref{evolution}) one immediately concludes
that the former are the function of eigenvalues of the Casimir operator
$\mbox{\boldmath${\cal J}$}{}^2$. Thus the interaction kernels can be
expressed as functions of Casimir operator in a given subchannel, e.g.\
$K_{ab} = h (\mbox{\boldmath${\cal J}$}{}^2_{ab})$.

Let us give an explicit example of a basis, convenient for present studies.
Choose a space ${\cal V} = \{ \theta^k | k = 0,1,\dots,\infty \}$ spanned
by elements $\theta^k \equiv \partial^k_+ \phi / \Gamma (k + \nu + 1)$.
In this representation $[{\cal J}^{\pm,3} ,\chi ( \theta )]_- =
\hat J^{\pm,3} \chi ( \theta )$ the generators take the following form
\begin{equation}
\hat J^+ = (\nu + 1) \theta + \theta^2 \partial_\theta ,
\quad
\hat J^- = \partial_\theta,
\quad
\hat J^3 = \ft12 (\nu + 1) + \theta \partial_\theta ,
\end{equation}
where $\partial_\theta \equiv \frac{\partial}{\partial\theta}$ and the
quadratic Casimir operator reads ${\mbox{\boldmath$\hat J$}}{}^2 = \hat J^3
( \hat J^3 - 1 ) - \hat J^+ \hat J^-$. The advantage of this basis lies in
the fact that conformal operator has the form of translation, so that the
highest weight vector condition is easier to solve. For a multi-variable
function $\chi (\theta) \equiv \chi ( \theta_1, \theta_2,\dots,\theta_n )$
the operators are defined by the sum of single particle ones as
$\hat J^{\pm,3} = \sum_{\ell = 1}^{n} \hat J^{\pm,3}_\ell$ and they obey
the usual commutation rules $[\hat J^3 , \hat J^\pm]_- = \pm \hat
J^\pm$, $[\hat J^+ , \hat J^-]_- = -2 \hat J^3$. Obviously, a single
particle state $\theta_\ell$ is an eigenstate of the Casimir operator
${\mbox{\boldmath$\hat J$}}{}^2_a \theta_a = {\mbox{\boldmath$\hat
\jmath$}}{}^2_a \theta_a$, with ${\mbox{\boldmath$\hat \jmath$}}{}^2_a
\equiv \ft14 ( \nu_a^2 - 1 )$. The eigenstates of two-particle Casimir
operator ${\mbox{\boldmath$\hat J$}}{}^2_{ab}$ in $(ab)$-subchannel are
$( {\hat J}^+_{ab} )^k \theta_{ab}^j$ (here and throughout $\theta_{ab}
\equiv \theta_a - \theta_b$) and coincide with the bilinear conformal
operator (\ref{ConfOper}) and possess the same eigenvalues. Now, the
three-particle basis efficient for present applications has to be chosen
so that it diagonalizes the total three-particle Casimir operator
${\mbox{\boldmath$\hat J$}}{}^2 = {\mbox{\boldmath$\hat J$}}{}^2_{12}
+ {\mbox{\boldmath$\hat J$}}{}^2_{23} + {\mbox{\boldmath$\hat J$}}{}^2_{13}
- \sum{}_{\ell = 1}^{3} {\mbox{\boldmath$\hat \jmath$}}{}^2_\ell $ and
one in a subchannel ${\mbox{\boldmath$\hat J$}}{}_{ab}^2$, say $a = 1$
and $b = 2$,
\begin{eqnarray}
\label{DiagTot}
&&{\mbox{\boldmath$\hat J$}}{}^2
{\cal P}_{J; j}
= \left( J + \ft12 (\nu_1 + \nu_2 + \nu_3 + 3 )\right)
\left( J + \ft12 (\nu_1 + \nu_2 + \nu_3 + 1 )\right)
{\cal P}_{J; j} , \nonumber\\
\label{Diag12}
&&{\mbox{\boldmath$\hat J$}}{}^2_{12}
{\cal P}_{J; j}
= \left( j + \ft12 ( \nu_1 + \nu_2) \right)
\left( j + \ft12 ( \nu_1 + \nu_2 + 2 ) \right)
{\cal P}_{J; j} .
\end{eqnarray}
The solution is given in terms of hypergeometric function
${\cal P}_{J;j} (\theta_1, \theta_2 | \theta_3) \propto \theta_{12}^J
\theta^{j - J} {_2\!F_1} (j - J, j + \nu_1 + 1, 2j + \nu_1 + \nu_2 + 2
| \theta)$ with $\theta \equiv \theta_{12}/\theta_{32}$.

Since the physical anomalous dimensions have to be real we have to define
an appropriate scalar product resulting in selfadjoint Hamiltonian,
namely
\begin{equation}
\label{ScalarProduct}
\langle \chi ( \theta ) | \chi ( \theta ) \rangle
= \int_{\bigcup^n_{\ell = 1} \{ |\theta_\ell| \leq 1 \}}
\prod_{\ell = 1}^{n}
\frac{d \theta_\ell d \bar\theta_\ell}{2 \pi i}
( 1 - \theta_\ell \bar\theta_\ell )^{\nu_\ell - 1}
\chi ( \bar\theta ) \chi ( \theta ) ,
\end{equation}
and $\bar\theta = \theta^\ast$. Then it can be seen that the Casimir
operator and, as a consequence, the Hamiltonian are selfadjoint operators
w.r.t.\ such defined inner product $( {\mbox{\boldmath$\hat J$}}{}^2
)^\dagger = {\mbox{\boldmath$\hat J$}}{}^2$, $h^\dagger = h$, and
therefore ${\rm Im}\, {\cal E} = 0$.

Obviously, to solve the three particle problem (\ref{ThreePartProblem}) one
has to find an extra integral of motion in addition to the one provided by
the conformal symmetry. As we show in the next sections it turns out that
this is the case of antiquark-gluon-quark and three-gluon systems in some
limits.

%%%%%%%%%%%%%%%%%%%%%%%%%%%%%%%%%%%%%%%%%%%%%%%%%%%%%%%%%%%%%%%%%%%%%
\section{Antiquark-gluon-quark correlators.}
\label{qGq}
%%%%%%%%%%%%%%%%%%%%%%%%%%%%%%%%%%%%%%%%%%%%%%%%%%%%%%%%%%%%%%%%%%%%%

The antiquark-gluon-quark correlations contribute to a number of single
spin asymmeties \cite{SinSip85} and enter as a three-particle part in the
twist-3 (skewed) parton distributions \cite{BukKurLip84,JafJi92,BelMul97}.
For instance, the transverse spin structure function $g_2$ admits
the following three-parton piece
\begin{equation}
g_2^{\rm tw-3} (x)
= \int d x_1 d x_3 \, {\cal J} (x_1, x_3) \, Y (x_1, x_3) ,
\end{equation}
where ${\cal J}$ is a differential operator acting on the momentum
fractions $x_i$ of a two-argument function $Y (x_1, x_3)$ which is
defined as a light-cone Fourier transform of a hadron matrix element
of a $C$-even combination $\langle p | {\cal S}^+ (\kappa_1, \kappa_3)
+ {\cal S}^- (- \kappa_3, - \kappa_1) | p \rangle$ of nonlocal operators
${\cal S}^\pm_\mu (\kappa_1, \kappa_3) = i \g \bar\psi ( \kappa_3 n )
\gamma_+ [ i \widetilde G^\perp_{\mu +} (0) \pm \gamma_5 G^\perp_{\mu +}
(0) ] \psi ( \kappa_1 n )$. This will be the case of study in this
section \cite{BraDerMan98,BelGGG,DerKorMan99}, in particular we elaborate
on operator ${\cal S}^+$, since ${\cal S}^-$ is related to it by charge
conjugation. Similar considerations apply to chiral odd sector
\cite{BraDerMan98,BelQGQ,DerKorMan99}.

In view of our discussion in Section \ref{evolution} the total
$\mbox{\boldmath$K$}_{\bar q g q}$ evolution kernel splits into a sum
$K_{\bar q g} + K_{g q} + K_{\bar q q} - \ft14 \beta_0$, where the
$\beta$-function term is due to presence of the coupling constant in the
definition of the operators ${\cal S}$. The leading twist QCD interaction
kernels $K_{ab}$ with non-zero momentum transfer in $t$-channel are known
for a long time, see e.g.\ \cite{BukFroLipKur85}. One of the complications
to solve the three-particle system is due to non-trivial colour structures
of the latter, e.g.\ $K_{\bar q q}$ depends on $C_F - \ft12 C_A$, while
$K_{(\bar q, q) g}$ on both $C_F - \ft12 C_A$ and $C_F$. Obviously, the
kernel simplifies drastically in multicolour limit $N \to \infty$,
since the interaction of end point quarks effectively vanishes, $C_F - \ft12
C_A \sim {\cal O} (N_c^{-1})$. In this case the total interaction kernel
takes a form $\mbox{\boldmath$K$}_{\bar q g q} = \ft12 N_c {\cal H}$ where
\begin{equation}
\label{IntegHamil}
{\cal H} = h_{\bar q g} \left( \ft32 \right) + h_{g q} \left( \ft12 \right)
- \ft32,
\quad\mbox{with}\quad
h_{ab} (\delta) = \psi \big( {\hat J}_{ab} + \delta \big)
+ \psi \big( {\hat J}_{ab} - \delta \big) - 2 \psi (1) ,
\end{equation}
and ${\mbox{\boldmath$\hat J$}}{}^2 = \hat J (\hat J - 1)$. Thus we
have reduced the original problem to an open chain of three particles on
a line with conformal invariant interaction of neighbors (quarks and
gluons) and no interaction of end points (quarks). This observation turns
out to be fruitful since one can find the pair-wise Hamiltonians
(\ref{IntegHamil}) among the series of integrable ones for
inhomogeneous spin chains. It can be easily generated from
the Yang-Baxter bundle \cite{KulResSkl81}, which is a solution to the
Yang-Baxter equation \cite{KorBogIze93},
\begin{equation}
\label{YBbundle}
R_{ab} (\lambda) = f(\nu, \lambda) P_{ab}
\frac{\Gamma (\hat J_{ab} + \lambda) \Gamma (\nu + 1 - \lambda)}{
\Gamma (\hat J_{ab} - \lambda) \Gamma (\nu + 1 + \lambda)} ,
\end{equation}
as follows $h_{ab} (\delta) = R_{ab} (- \delta) R'_{ab} (- \delta)$,
provided the function $f$ is chosen like this $f (\nu, \lambda) =
\frac{\Gamma (1 - \lambda)}{\Gamma (1 + \lambda)} \frac{\Gamma
(\nu + 1 + \lambda)}{\Gamma (\nu + 1 - \lambda)}$. The total ${\cal H}$
can be produced from the open spin chain transfer matrix \cite{Skl88}
$t_b (\lambda) = {\rm tr}_b \left. T_b (\lambda) T_b^{- 1} (- \lambda)
\right|_{\nu_b = 2}$ constructed out of $R$'s, $T_b (\lambda) =
R_{a_1 b} (\lambda - \delta_1) R_{a_2 b} (\lambda) R_{a_3 b} (\lambda
- \delta_3)$ with the spins of $a_i$ spaces chosen like $\nu_1 =
\nu_3 = 1$, $\nu_2 = 2$, using the formula ${\cal H} = \ft12
\ft{d}{d \lambda} \ln t_b (\lambda)$. Identifying imhomogeneities as
$\delta_1 = \ft32$ and $\delta_3 = \ft12$ we immediately find Eq.\
(\ref{IntegHamil}). This tells us that the antiquark-gluon-quark system
is exactly integrable in multicolour limit. This fact implies that there
exist a family of commuting integrals of motion. The first of
them is provided by the conformal symmetry of interaction and is given by
Casimir operator $\mbox{\boldmath$\hat J$}{}^2$. Thus, we have to find
only one `hidden' conserved charge. This follows from the standard
formalism of integrable spin chain models and forces us to calculate
transfer matrix with auxiliary space being two-dimensional. With
inhomogeneity parameters taken as above it gives an expansion in
rapidity $\lambda$ \cite{BelQGQ,BelGGG,DerKorMan99}
\begin{equation}
t_{\frac{1}{2}} (\lambda) = {\mit\Omega} (\lambda)
- ( 4 \lambda^2 - 1 )
( \lambda^2 - {\mbox{\boldmath$\hat \jmath$}}{}^2_2 )
\mbox{\boldmath$\hat J$}{}^2
- \ft12 ( 4 \lambda^2 - 1 )
\mbox{\boldmath${\cal Q}$} \left( \delta_1 , \delta_3 \right) ,
\end{equation}
with constant function ${\mit\Omega}$ and the `hidden' charge
\begin{equation}
\mbox{\boldmath${\cal Q}$} \left( \delta_1 , \delta_3 \right)
= [ {\mbox{\boldmath$\hat J$}}{}^2_{12},
{\mbox{\boldmath$\hat J$}}{}^2_{23} ]_+
- 2 \delta^2_1 {\mbox{\boldmath$\hat J$}}{}^2_{23}
- 2 \delta^2_3 {\mbox{\boldmath$\hat J$}}{}^2_{12} .
\end{equation}
Below we will find eigenfunctions of this much more simple (compared to
original Hamiltonian) operator $\mbox{\boldmath${\cal Q}$}$ and use them
to find the eigenvalues of the Hamiltonian (\ref{IntegHamil}). Let us
note in passing that integrable open spin chains have been encountered
as well in the solution of BFKL-type equation for quark-gluon reggeons
\cite{KarKir99}.

To find the eigenfunctions of the Hamiltonian (\ref{IntegHamil}) we
solve the equation for the charge with eigenfunctions ${\mit\Psi}$
expanded in the conformal basis discussed in Section \ref{conformal}
\begin{equation}
\label{RC1}
\mbox{\boldmath${\cal Q}$} \left( \ft32 , \ft12 \right) {\mit\Psi}
= q_S {\mit\Psi} ,
\qquad\qquad
{\mit\Psi} = \sum_{j = 0}^{J} \varrho_j {\mit\Upsilon}_j
{\cal P}_{J; j} (\theta_1, \theta_2 | \theta_3) ,
\end{equation}
with $\varrho_j^{- 1} = \left[(J - j + 1) (J + j + 5) (j + 1)^3 (j + 3)^3
/ (j + 2)^3 \right]^{1/2}$ being normalization coefficients. The main
advantage of the basis is a three-diagonal form of the matrix elements
of non-proper two-particle Casimir operators
$\mbox{\boldmath${\hat J}$}_{(1,2)3}$. Then one establishes that
Eq.\ (\ref{RC1}) is equivalent to the three-term recursion relation
\begin{equation}
\label{MasterEq}
(2j + 3) {\mit\Upsilon}_{j + 1}
+
(2j + 5) {\mit\Upsilon}_{j - 1}
+
\varrho_j^2\, (2j + 3)(2j + 5)
\left(
{[ \mbox{\boldmath${\cal Q}$} \left( \ft32 , \ft12 \right) ]}_{j,j}
- q_S
\right) {\mit\Upsilon}_j
= 0 ,
\end{equation}
where ${[ \mbox{\boldmath${\cal Q}$} ]}_{j,j}$ are the diagonal
elements of the charge $\langle {\cal P}_{J;j} | \mbox{\boldmath${\cal Q}$}
| {\cal P}_{J;j} \rangle$ in the basis ${\cal P}_{J; j} (\theta_1,
\theta_2 | \theta_3)$. Polynomiality requires the expansion
coefficients, ${\mit\Upsilon}_j$, to satisfy the boundary conditions
${\mit\Upsilon}_{- 1} = {\mit\Upsilon}_{J + 1} = 0$.

\begin{figure}[t]
\begin{center}
\hspace{0cm}
\mbox{
\begin{picture}(0,142)(100,0)
\put(-50,3){\insertfig{4.8}{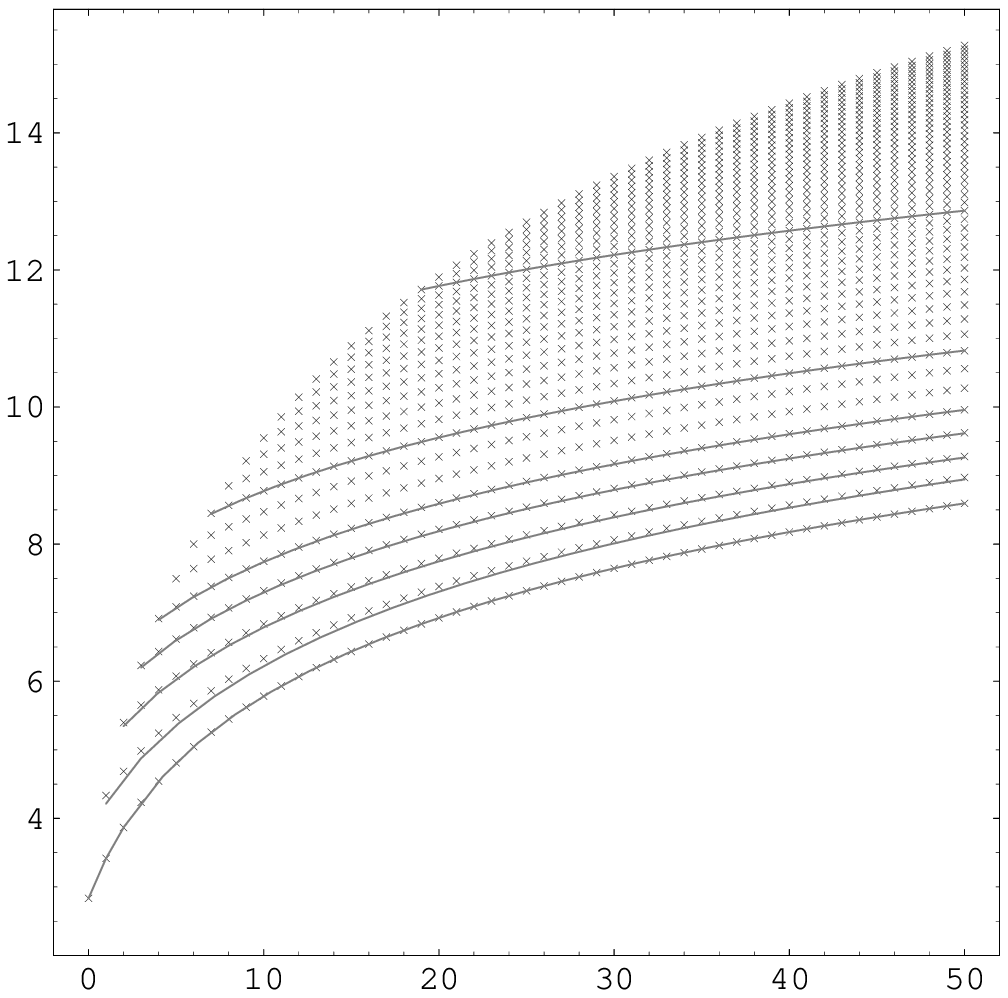}}
\put(110,0){\insertfig{5}{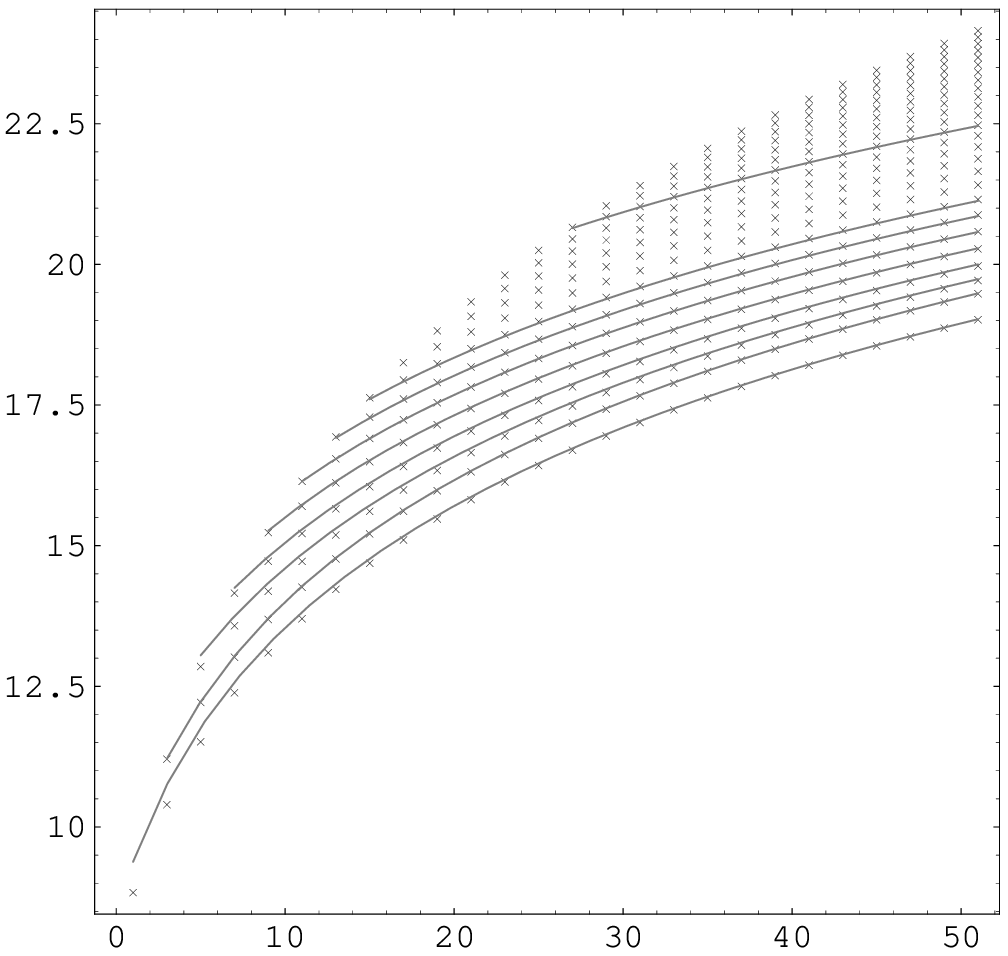}}
\put(-60,67){${\cal E}$}
\put(-30,120){$\bar q G q$}
\put(18,-5){$J$}
\put(104,67){${\cal E}$}
\put(136,120){$G \widetilde{G} G$}
\put(184,-5){$J$}
\end{picture}
}
\end{center}
\caption{\label{EnergySpectra} The energy of the systems ${\cal E} (J)$
versus $J$. (Left) The analytical trajectories from the Eqs.\
(\ref{LowestEnergy},\ref{TopAfterGap}) and a numerically diagonalized
anomalous dimension matrix. (Right) Same for gluons with analytical
result (\ref{EnerQ2app}). }
\end{figure}

The knowledge of eigenfunctions allows to find the energy of the
antiquark-gluon-quark system via the formula
\begin{equation}
\label{EnergyAnalyticLinear}
{\cal E} (J, q_S) =
\bigg(
\sum_{j = 0}^{J} (- 1)^j \frac{(j + 2)^3}{(j + 1)(j + 3)}
{\mit\Upsilon}_j
\bigg)^{-1}
\sum_{j = 0}^{J} (- 1)^j \epsilon (j)
\frac{(j + 2)^3}{(j + 1)(j + 3)} {\mit\Upsilon}_j
+ \ft13 ,
\end{equation}
with $\epsilon (j) = \psi (j + 1) + \psi (j + 4) - 2 \psi (1)$.
Unfortunately, the exact solution to (\ref{MasterEq}), and thus
(\ref{EnergyAnalyticLinear}), can hardly be found. Therefore, we restrict
ourselves to WKB approximation for large conformal spins $J$. For the
levels behaving as ${\cal E} \propto 2 \ln J$ we can find the
exact\footnote{For the chiral odd distribution similar result has been
found in Refs.\ \cite{AliBraHil91,BelMul97} and for twist-three
fragmentation functions in \cite{Bel97a}.} lowest trajectory
\cite{AliBraHil91}
\begin{equation}
\label{LowestEnergy}
{\cal E} (J) = \psi (J + 3) + \psi (J + 4) - 2 \psi (1) - \ft12 .
\end{equation}
The remainder of the spectrum is described by the formula
\cite{BraDerMan98,BelGGG,DerKorMan99}
\begin{equation}
\label{TopAfterGap}
{\cal E} (J, q_S)
= 2 \ln J - 4 \psi (1)
+ 2 \, {\rm Re} \, \psi \left( \ft32 + i \eta_S \right)
- \ft32 ,
\end{equation}
with $\eta_S = \ft12 \sqrt{2 q_S/J^2 - 3}$. The conserved charge is
quantized, with WKB quantization condition arisen from the matching of the
WKB solution with exact ones at `turning' points, and gives quantized
values of energy via (\ref{TopAfterGap}). It compares well with an
explicit numerical diagonalization of the anomalous dimension matrix
\cite{BukKurLip84} as shown in Figs.\ \ref{EnergySpectra}. There is an
alternative description of the spectrum by trajectories which behave
as ${\cal E} \propto 4 \ln J$ at large $J$. The dependence on the
conserved charge reads asymptotically ${\cal E} (J, q_S) = \ln \ft12 q_S
- 4 \psi (1) - \ft32 + {\cal O} (J^{- 1})$. The first few expansion
coefficients of energy in the series in $J^{-1}$ was found in
\cite{BelGGG,DerKorMan99}. Corrections in $N_c^{-1}$ have been discussed
recently in Ref.\ \cite{BraKorMan00}.

%%%%%%%%%%%%%%%%%%%%%%%%%%%%%%%%%%%%%%%%%%%%%%%%%%%%%%%%%%%%%%%%%%%%%
%                           Figure 2                                %
%%%%%%%%%%%%%%%%%%%%%%%%%%%%%%%%%%%%%%%%%%%%%%%%%%%%%%%%%%%%%%%%%%%%%
%%%%%%%%%%%%%%%%%%%%%%%%%%%%%%%%%%%%%%%%%%%%%%%%%%%%%%%%%%%%%%%%%%%%%

%%%%%%%%%%%%%%%%%%%%%%%%%%%%%%%%%%%%%%%%%%%%%%%%%%%%%%%%%%%%%%%%%%%%%
\section{Three-gluon correlators.}
\label{GGG}
%%%%%%%%%%%%%%%%%%%%%%%%%%%%%%%%%%%%%%%%%%%%%%%%%%%%%%%%%%%%%%%%%%%%%

\begin{figure}[t]
\begin{center}
\hspace{0cm}
\mbox{
\begin{picture}(0,45)(100,0)
\put(45,-5){\insertfig{5}{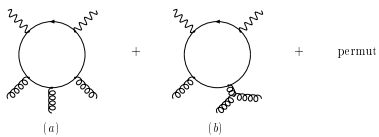}}
\end{picture}
}
\end{center}
\caption{\label{bubbles} Coefficient function of three-gluon state
in the structure function $g_2$.}
\end{figure}

Three-gluon correlators enter the $g_2$ via the quark loop coupling
shown in Fig.\ \ref{bubbles} and might be responsible for its
small-$x$ behaviour,
\begin{equation}
g_2 (x) = \int d x_1 d x_3 {\cal C}_{g \tilde g g} (x_1, x_3)
\, G (x_1, x_3) .
\end{equation}
Here ${\cal C}_{g \tilde g g} = {\cal C}_{g \tilde g g}^{(a)} +
{\cal C}_{g \tilde g g}^{(b)}$, with ${\cal C}_{g \tilde g g}^{(a)}$
calculated in Ref.\ \cite{BelEfrTer95} but contact-type contributions $(b)$
not accounted for. The function $G (x_1, x_2)$ is a Fourier transform of
non-local operators, e.g.\ ${\cal G}_\mu (\kappa_1, \kappa_3) =
\g f_{abc}\, G^{a \perp}_{+ \nu} ( \kappa_3 n )
\widetilde{G}^{b \perp}_{+ \mu} (0) G^{c \perp}_{+ \nu} ( \kappa_1 n )$.

The analysis, equivalent to the one done for the antiquark-gluon-quark
correlator, gives the evolution kernel $\mbox{\boldmath$K$}_{g \tilde g g}$
which can be expressed up to corrections playing negligible role in the
generation of the spectrum of anomalous dimensions as a sum
$\mbox{\boldmath$K$}_{g \tilde g g} = \ft12 C_A {\cal H} + \ft12 \beta_0$
where ${\cal H}$ is split into two pieces,
\begin{equation}
\label{PertHamilton}
{\cal H} = {\cal H}_0 + {\cal V},
\qquad\mbox{where}\qquad
{\cal H}_0 = h_{12} + h_{23} + h_{31},
\qquad
{\cal V} = v_{12} + v_{23},
\end{equation}
and pair-wise interactions given by
\begin{equation}
\label{QCDlocalHam}
h_{ab} = 2 \psi \big( \hat J_{ab} \big) - 2 \psi (1) ,
\qquad
v_{ab} = - 4 \, \mbox{\boldmath$\hat J$}{}_{ab}^{-2} .
\end{equation}
Obviously, the separation into ${\cal H}_0$ and ${\cal V}$ is not
accidental, but reflect the fact that ${\cal H}_0$, which describes
the gluon interaction with aligned helicities, coincides with the
Hamiltonian of XXX closed magnet of noncompact spin $-\ft32$. This can
be easily found by means of the same $R$ matrix (\ref{YBbundle})
from the logarithmic derivative, ${\cal H}_0 = \ft{d}{d\lambda}\ln
t_b (\lambda)$, of the transfer matrix of an open spin chain $t_b (\lambda)
= {\rm tr}_b R_{a_1 b} (\lambda) R_{a_2 b} (\lambda) R_{a_3 b} (\lambda)$.
As a result the system described by ${\cal H}_0$ contains a family
of mutually commuting conserved charges which arise from the expansion
of the auxiliary transfer matrix
\begin{equation}
t_\frac{1}{2} (\lambda)
= 2 \lambda^3 + \lambda \left( {\mbox{\boldmath$\hat J$}}^2
- 3 {\mbox{\boldmath$\hat \jmath$}}^2 \right)
+ i \mbox{\boldmath${\cal Q}$},
\qquad\mbox{with}\qquad
\mbox{\boldmath${\cal Q}$} = \ft{i}2 [ {\mbox{\boldmath$\hat J$}}^2_{12} ,
{\mbox{\boldmath$\hat J$}}^2_{23} ]_- .
\end{equation}
Note that the same Hamiltonian, however of different conformal
spin, has been encountered in interaction of QCD reggeons
\cite{Lip94,FadKor95} and Brodsky-Lepage evolution equation
for baryons \cite{BraDerKorMan99} in the case of the same helicities
of participating quarks. The coincidence of quark and gluon kernels
for the same helicity states is a mere consequence of ${\cal N} = 1$
supersymmetry which relates corresponding kernels \cite{BelMul99},
$h^T_{\bar q q} ( 2n + 1 ) = h^T_{g g} (2n + 1) = h^T_{q g} (2n + 1)$
where $h^T_{q g} (J) = 2 \psi (J + 2) + 2 \psi (J + 3) - 4 \psi (1) +
2 (- 1)^J / (J + 2) - 3$.

%%%%%%%%%%%%%%%%%%%%%%%%%%%%%%%%%%%%%%%%%%%%%%%%%%%%%%%%%%%%%%%%%%%%%
%                           Figure 3                                %
%%%%%%%%%%%%%%%%%%%%%%%%%%%%%%%%%%%%%%%%%%%%%%%%%%%%%%%%%%%%%%%%%%%%%
%%%%%%%%%%%%%%%%%%%%%%%%%%%%%%%%%%%%%%%%%%%%%%%%%%%%%%%%%%%%%%%%%%%%%

So our strategy will be, first, to diagonalize the exactly solvable
interaction and then to consider the effect of ${\cal V}$ interaction
on the spectrum. Thus in parallel to previous section, we
diagonalize first the charge $\mbox{\boldmath${\cal Q}$}$ and find
its eigenfunctions ${\mit\Upsilon}_j$ (\ref{RC1}), now
$\varrho_j^{-1} \equiv [ (j + 1)_4 (J + j + 6)_2 (J - j + 1)_2/
(2j + 5) ]^{1/2}$, from the recursion relation
\begin{equation}
\label{RecRel}
2 (2j + 5) \, q {\mit\Upsilon}_j
= a_j {\mit\Upsilon}_{j + 1} + a_{j + 1} {\mit\Upsilon}_{j - 1},
\end{equation}
with coefficients $a_j = (j + 1)(j + 3)(J + j + 6) (J - j + 2)$.
${\mit\Upsilon}$'s allow to find the energy via the formula
\begin{equation}
\label{EnegyLinear}
{\cal E}_0 (J, q) = 2 {\rm Re}\,
\frac{\sum_{j = 0}^{J} (- i)^j (2 j + 5) \epsilon (j) {\mit\Upsilon}_j}{
\sum_{j = 0}^{J} (- i)^j (2 j + 5) {\mit\Upsilon}_j} + 3 ,
\end{equation}
with the two-particle interaction $\epsilon (j) = 2 \psi (j + 3) -
2 \psi (1)$. The lowest trajectory can be found exactly
\begin{equation}
\label{Qoenergy}
{\cal E}_0 (J,0) = 2 \psi \left( \ft12 J + 3 \right)
+ 2 \psi \left( \ft12 J + 2 \right) - 4 \psi (1) + 4 .
\end{equation}
While the rest of the spectrum is described by WKB formula
\begin{equation}
\label{EnergyBottom}
{\cal E}_0 (J, q) = 4 \ln \eta - 6 \psi (1)
+ 2 {\rm Re}\, \psi \left( \ft32 - i q^\star \right) ,
\end{equation}
where $q^\star \equiv q/\eta^2$ and is valid with ${\cal O} (\eta^{-1})$.
We use the convention $\eta^2 \equiv \left( J + \ft32 (\nu + 1) \right)
\left( J + \ft32 (\nu + 1) - 1 \right)$ for the eigenvalues of the
quadratic Casimir operator of the chain which to the stated accuracy
reads for gluons $\eta = J + 4 + {\cal O} (J^{- 1})$. Matching of the
WKB and exact solutions to (\ref{RecRel}) gives a quantization conditions
for the charge, $q^\star \ln \eta = \arg \, \Gamma \left( \ft32 + i
q^\star \right) + \frac{\pi}{6} \left( 2 m + \ft12 [ 1 - (- 1)^J ]
\right)$. An alternative set of trajectories behaves like ${\cal E} (J, q)
= 2 \ln q - 6 \psi (1) + {\cal O} (J^{-1})$.

The perturbation ${\cal V}$ does not affect a bulk of the spectrum
except for the lowest few levels. To analyze the situation one considers
Hamiltonian ${\cal H}_\alpha = {\cal H} + \alpha {\cal V}$ specified by a
coupling constant $\alpha$ and study \cite{BraDerKorMan99,BelGGG} the
level flow as a function of $\alpha$. Making use of these results we
design the following formulae which describe well the spectrum of the
perturbed Hamiltonian, see Fig.\ \ref{EnergySpectra},
\begin{eqnarray}
\label{EnerQ2app}
{\cal E} (J, m) &=&
{\cal E}_0 (J, 0) ( \delta_{m,0} + \delta_{m,1})
+ {\cal E}_0 (J, q(m)) \theta (m - 2) - \Delta (m) .
\end{eqnarray}
Here ${\cal E}_0 (J, 0)$ is the ground state energy (\ref{Qoenergy})
and ${\cal E}_0 (J, q(m))$ is Eq.\ (\ref{EnergyBottom}) with the
$q (m)$ trajectories deduced from the quantization condition. Finally,
\begin{eqnarray}
\Delta (m) = 0.54 \, \delta_{m,0} + 0.08 \, \delta_{m,1}
+ \left( \delta_0 - \delta (m - 2) \right)
\theta \left( \delta_0 - \delta (m - 2) \right),
\end{eqnarray}
is the shift-function with $\delta_0 = 0.15$ and $\delta = 0.01$.

%%%%%%%%%%%%%%%%%%%%%%%%%%%%%%%%%%%%%%%%%%%%%%%%%%%%%%%%%%%%%%%%%%%%%
\section{Conclusions}
%%%%%%%%%%%%%%%%%%%%%%%%%%%%%%%%%%%%%%%%%%%%%%%%%%%%%%%%%%%%%%%%%%%%%

To conclude, it remains a challenge for QCD to understand the
appearance of integrable structures for light-cone
\cite{BraDerMan98,BraDerKorMan99,BelQGQ,BelGGG,DerKorMan99} and
transverse momentum \cite{Lip94,FadKor95,KarKir99} dynamics and
unravel their striking similarity.

%%%%%%%%%%%%%%%%%%%%%%%%%%%%%%%%%%%%%%%%%%%%%%%%%%%%%%%%%%%%%%%%%%%%%
\section*{Acknowledgments}
%%%%%%%%%%%%%%%%%%%%%%%%%%%%%%%%%%%%%%%%%%%%%%%%%%%%%%%%%%%%%%%%%%%%%

We would like to thank G.P. Korchemsky for discussions and A. Sch\"afer
for the hospitality at the Institut f\"ur Theoretische Physik,
Universit\"at Regensburg. This work was supported by Alexander von
Humboldt Foundation and in part by National Science Foundation, under
grant PHY9722101.

%%%%%%%%%%%%%%%%%%%%%%%%%%%%%%%%%%%%%%%%%%%%%%%%%%%%%%%%%%%%%%%%%%%%%
\section*{References}
%%%%%%%%%%%%%%%%%%%%%%%%%%%%%%%%%%%%%%%%%%%%%%%%%%%%%%%%%%%%%%%%%%%%%

\end{document}